\documentclass[aip,pop,reprint,groupedaddress]{revtex4-1}
\usepackage{bm}
\usepackage{amsmath}
\usepackage[allcolors=blue,colorlinks=true]{hyperref}
\allowdisplaybreaks
\begin{document}
\title[A reanalysis of a strong-flow gyrokinetic formalism]{A reanalysis of a strong-flow gyrokinetic formalism}
\author{A. Y. Sharma}
\author{B. F. McMillan}
\affiliation{Centre for Fusion, Space and Astrophysics, Physics Department, University of Warwick, UK}
\date{\today}
\begin{abstract}
We reanalyse an arbitrary-wavelength gyrokinetic formalism [A. M. Dimits, Phys. Plasmas {\bf17}, 055901 (2010)], which orders only the vorticity to be small and allows strong, time-varying flows on medium and long wavelengths. We obtain a simpler gyrocentre Lagrangian up to second order. In addition, the gyrokinetic Poisson equation, derived either via variation of the system Lagrangian or explicit density calculation, is consistent with that of the weak-flow gyrokinetic formalism [T. S. Hahm, Phys. Fluids {\bf31}, 2670 (1988)] at all wavelengths in the weak flow limit. The reanalysed formalism has been numerically implemented as a particle-in-cell code. An iterative scheme is described which allows for numerical solution of this system of equations, given the implicit dependence of the Euler-Lagrange equations on the time derivative of the potential.
\end{abstract}
\maketitle

\section{Introduction}

The weak-flow gyrokinetic formalism\cite{Hahm1988,Dimits1992} uses a gyrokinetic ordering parameter
\begin{equation}
\epsilon\sim\omega/\Omega\sim v_{\rm E\times B}/v_{\rm t} \ll 1,  \label{o1}
\end{equation}
with $\omega$ a characteristic frequency, $\Omega$ the gyrofrequency, $v_{\rm E\times B}$ the $\rm E\times B$ drift speed and $v_{\rm t}$ the typical thermal speed. 

The ordering \eqref{o1} may be poorly satisfied in the core and edge of tokamaks because of either
large overall rotation or relatively strong flows in the tokamak pedestal. 
It is also frequently broken in 
astrophysical plasmas. Various approaches\cite{Hahm1996,Miyato2009} to including stronger flows
in a gyrokinetic framework have been proposed, but the most general so far\cite{Dimits2010} is based 
on ordering the vorticity to be small, 
\begin{equation}
\epsilon\sim v_{\rm E\times B}'/\Omega,
\label{os}
\end{equation}
where $v_{\rm E\times B}'$ is the characteristic magnitude of the spatial derivatives of the $\rm E\times B$ drift velocity. This is a maximal ordering in the sense that a larger vorticity on any scale would lead to breaking of
the magnetic moment invariance, as nonlinear frequencies are comparable to the vorticity. 
Ordering the vorticity allows for general large, time varying flows on large length scales
as well as gyroscale perturbations, and includes them within a single description, unlike schemes
based on separation of scales\cite{Qin2007,Brizard1995} or long-wavelength schemes\cite{Miyato2009}.

However, in the weak-flow limit, the gyrokinetic Poisson equation of Ref. \onlinecite{Dimits2010} disagrees with that of the 
weak-flow gyrokinetic formalism at wavelengths comparable to the gyroradius. We rederive this theory and explain some
minor but important departures from the derivation of the weak-flow theory. In our reanalysis, we obtain a Poisson equation, via both a variational and direct method, that, in the weak-flow limit, agrees with the weak-flow gyrokinetic Poisson equation at all wavelengths.

\section{Guiding-centre Lagrangian}

The particle fundamental 1-form for electrostatic perturbations in a slab uniform equilibrium magnetic field is\begin{equation}
\gamma=[\bm{A}(\bm{x})+\bm{v}]\cdot{\rm d}\bm{x}-\left[\tfrac{1}{2}\bm{v}^2+\phi(\bm{x},t)\right]{\rm d}t,\label{g}\end{equation}
where we use units such that $q=T=m=v_{\rm t}=1$, $q$ is the particle charge, $T$ is the temperature, $m$ is the particle mass, $\bm{A}$ is the magnetic vector potential, $\bm{x}$ is the particle position, $\bm{v}$ is the particle velocity and $t$ is time. We redefine $\bm{v}$ as the velocity in a frame moving with a velocity \(\bm{u}(\bm{x},\bm{v},t)\) such that Eq. \eqref{g} becomes\begin{equation}
\gamma=[\bm{A}(\bm{x})+\bm{v}+\bm{u}]\cdot{\rm d}\bm{x}-\left[\tfrac{1}{2}(\bm{v}+\bm{u})^2+\phi\right]{\rm d}t.\label{gt}\end{equation}
The guiding-centre fundamental 1-form (Appendix \ref{agc}) is\begin{equation}\begin{aligned}
\Gamma=&[\bm{A}(\bm{R})+U\hat{\bm{b}}+\bm{u}]\cdot{\rm d}\bm{R}-\bm{\rho}\cdot{\rm d}\bm{u}+\mu{\rm d}\theta\\&
-(\tfrac{1}{2}U^2+\mu\Omega+\tfrac{1}{2}\bm{u}^2+\langle\phi\rangle+\delta_1\tilde{\phi}){\rm d}t,\end{aligned}\label{gc}\end{equation}
\[\delta_1\tilde{\phi}=\tilde{\phi}+\bm{\rho}\cdot\bm{\Omega}\times\bm{u},\]
where $\bm{R}=\bm{x}-\bm{\rho}$ is the guiding-centre position, $\bm{\rho}=v_\perp\Omega^{-1}(\cos\theta\hat{1}-\sin\theta\hat{2})$ is the gyroradius, $v_\perp$ is the perpendicular speed, $\theta$ is the gyroangle defined with the opposite sign to that of Ref. \onlinecite{Dimits2010}, $\hat{1}=\hat{2}\times\hat{\bm{b}}$, $\hat{\bm{b}}$ is the magnetic field unit vector, $U=\hat{\bm{b}}\cdot\bm{v}$ is the parallel speed, $\mu=\tfrac{1}{2}v_\perp\Omega^{-1}$ is the magnetic moment, $\langle\dots\rangle\equiv(2\pi)^{-1}\oint{\rm d}\theta(\dots)$, $\tilde{\phi}=\phi-\langle\phi\rangle$, $\bm{\Omega}=\Omega\hat{\bm{b}}$ and we have used the gauge\[
S=-\bm{\rho}\cdot\big[\big(\tfrac{1}{2}\bm{\rho}\cdot\bm{\nabla}+1\big)\bm{A}(\bm{R})+\bm{u}\big].\]

\section{Gyrocentre Lagrangian}

Using the ordering \eqref{os}, $x\sim 1$ and
\begin{equation}
\bm{u}=\Omega^{-1}\hat{\bm{b}}\times\bm{\nabla}\langle\phi\rangle,\label{u}
\end{equation}
we can order the terms in the Lagrangian in terms of their variation over typical length scales as 
\begin{equation}
\Gamma  =  \Gamma_0+\Gamma_1,
\end{equation}
with 
\begin{equation}
 \Gamma_1= -\bm{\rho}\cdot{\rm d}\bm{u}-\delta_1\tilde{\phi}{\rm d}t.
\end{equation}
As in weak-flow formalisms, the lowest order Lagrangian $\Gamma_0$ contains terms which may be large on sufficiently long length scales.
As in Ref. \onlinecite{Dimits2010} and in addition to the conditions in Appendix \ref{agy}, $\bm{u}$ must satisfy the condition
\[\left(\partial_t+\bm{u}\cdot\bm{\nabla}\right)\bm{u}\sim\epsilon^2.\]
We use noncanonical Hamiltonian Lie-transform perturbation theory\cite{Littlejohn1982,Cary1983} to determine a set of gyrocentre coordinates where the Lagrangian is $\theta$-independent. 
This procedure systematically removes the $\theta$-dependence from the Lagrangian order by order. The transformation between guiding-centre and gyrocentre space is then
given in terms of a Lie transform of the form
\[\mathsf{T}^{\pm1}=\exp\left(\pm\sum_{n=1}\epsilon^n\mathcal{L}_n\right),\]
where $\mathcal{L}_n\Gamma=g^a_n\omega_{ab}{\rm d}Z^b$, $g^a_n$ are the generators, $a,b\in\{0,\dots,6\}$, \begin{equation}
\omega_{ab}=\Gamma_{b,a}-\Gamma_{a,b}\label{lm}\end{equation}
are the Lagrange matrix components and $\Gamma_{b,a}=\partial_a\Gamma_b$ (Einstein notation is used). The requirement that the first-order Lagrangian
be $\theta$-independent, with the choice  $g^t_n=0$, yields (Appendix \ref{agy}) the non-zero first-order generators
\begin{equation}
\begin{aligned}
  & g^{\bm{R}}_1  =\Omega^{-2}\bm{\nabla}\tilde{\Phi}\times\hat{\bm{b}},           \\
  & g^\mu_1            =\Omega^{-1}\delta_1\tilde{\phi},                                                   \\
  & g^\theta_1          =\bm{\rho}\cdot\bm{u}_{,\mu}-\Omega^{-1}\delta_1\tilde{\Phi}_{,\mu}
                      =-\Omega^{-1}\tilde{\Phi}_{,\mu}-\bm{u}\cdot\bm{\rho}_{,\mu},
\end{aligned}
\label{gs}
\end{equation}
where $\delta_1\tilde{\Phi}=\int{\rm d}\theta\delta_1\tilde{\phi}$ and $\tilde{\Phi}=\int{\rm d}\theta\tilde{\phi}$. Given a long wavelength flow, $g^\mu_1$ and $g^\theta_1$ are smaller in this strong flow formalism than in the equivalent 
weak-flow formalism, reflecting the improvement in the ordering scheme for such a case. 
Unlike Ref. \onlinecite{Dimits2010}, we simplify the second order Lagrangian by moving the second order terms into the time component (Appendix \ref{agy}). The gyrocentre Lagrangian up to second order is
\begin{equation}
\begin{aligned}
  \bar{\Gamma}=&[\bm{A}(\bar{\bm{R}})+\bar{U}\hat{\bm{b}}]\cdot{\rm d}\bar{\bm{R}}+\bar{\mu}{\rm d}\bar{\theta}-\big(\tfrac{1}{2}\bar{U}^2+\bar{\mu}\Omega+\langle\phi\rangle\\&
  -\tfrac{1}{2}\langle g^{\bar{\bm{R}}}_1\cdot\bar{\bm{\nabla}}\tilde{\phi}\rangle-\tfrac{1}{2}\Omega^{-1}\langle\tilde{\phi}^2\rangle_{,\bar{\mu}}\big){\rm d}t+\bar{\bm{u}}\cdot\left({\rm d}\bar{\bm{R}}-\bar{\bm{u}}{\rm d}t\right),
\end{aligned}
\label{gy}
\end{equation}
where the overbar denotes a gyrocentre quantity. The last term is the only one absent from the weak-flow gyrocentre Lagrangian at this order; the main qualitative difference
with the weak-flow formalism is simply the presence of the electric potential in the symplectic part of the Lagrangian.
\section{Euler-Lagrange equations}

Using the gyrocentre Lagrangian up to first order, the Euler-Lagrange equations in terms of gyrocentre coordinates,
\[\bar{\omega}_{ij}\dot{\bar{Z}}_j=\bar{\omega}_{ti},\]
where $i,j\in\{1,\dots,6\}$, yield (Appendix \ref{ael})\begin{equation}\begin{aligned}
&\dot{\bar{\bm{R}}}=\bar{\bm{u}}+\bar{\Omega}^{*-1}_\|\hat{\bm{b}}\times\left(\partial_t+\bar{\bm{u}}\cdot\bar{\bm{\nabla}}+\bar{U}\bar{\nabla}_\|\right)\bar{\bm{u}}+\bar{U}\hat{\bm{b}},\\&
\dot{\bar{U}}=-\langle\phi\rangle_{,\bar{z}}+\bar{\Omega}^{*-1}_\|\bar{\bm{u}}_{,\bar{z}}\cdot\hat{\bm{b}}\times\left(\partial_t+\bar{\bm{u}}\cdot\bar{\bm{\nabla}}\right)\bar{\bm{u}},\\&
\dot{\bar{\mu}}=0,\\&
\dot{\bar{\theta}}=\Omega+\langle\phi\rangle_{,\bar{\mu}}-\bar{\Omega}^{*-1}_\|\bar{\bm{u}}_{,\bar{\mu}}\cdot\hat{\bm{b}}\times\left(\partial_t+\bar{\bm{u}}\cdot\bar{\bm{\nabla}}+\bar{U}\bar{\nabla}_\|\right)\bar{\bm{u}},\\&
\bar{\Omega}^*_\|=\Omega+\hat{\bm{b}}\cdot\bar{\bm{\nabla}}\times\bar{\bm{u}}.\end{aligned}\label{el}\end{equation}
Note that we recover an additional term in the $\dot{\bar{U}}$ equation which appears to be missing in Ref. \onlinecite{Dimits2010}. Physically, 
it is a ponderomotive term that typically results from the appearance of a $\bar{u}^2$ term in the Lagrangian\cite{Krommes2013}; the analogue of this term is present in Ref. \onlinecite{Hahm1996}. 
The contributions to the equations of motion from the second order part of the Lagrangian are\begin{align*}&
\dot{\bar{\bm{R}}}_2=-\bar{\Omega}^{*-1}_\|\hat{\bm{b}}\times\bar{\bm{\nabla}}\bar{H}_2,\\&
\dot{\bar{U}}_2=\bar{H}_{2,\bar{z}}-\bar{\Omega}^{*-1}_\|\bar{\bm{u}}_{,\bar{z}}\cdot\hat{\bm{b}}\times\bar{\bm{\nabla}}\bar{H}_2,\\&
\dot{\bar{\theta}}_2=-\bar{H}_{2,\bar{\mu}},\\&
\bar{H}_2=\tfrac{1}{2}\langle g^{\bar{\bm{R}}}_1\cdot\bar{\bm{\nabla}}\tilde{\phi}\rangle+\tfrac{1}{2}\Omega^{-1}\langle\delta_1\tilde{\phi}^2\rangle_{,\bar{\mu}}+\hat{\bm{b}}\times\langle\delta_1\tilde{\phi}\bar{\bm{\rho}}\rangle\cdot\bar{\bm{u}}_{,\bar{\mu}},\end{align*}
where $\bar{H}_2$ is the second order part of the gyrocentre Hamiltonian. The equations of motion that include the contributions from the second order part of the Lagrangian can be simplified by renormalising the potential\cite{Lee1983}.

\section{Poisson equation}
Gyrokinetic Poisson and Amp\`{e}re's equations have previously been obtained by varying the system Lagrangian with respect to the field variables
\cite{Scott2010,Brizard2007}. We find it helpful to give an elementary explanation of why this should be possible.

First, consider the many-body Lagrangian for a set of point particles interacting with a field, with integral terms for the field self-interaction:
this is a well posed problem at least if we restrict the fields to be sufficiently smooth, and Euler-Lagrange equations for the particles and
the usual Maxwell equations are directly obtained by varying particle coordinates and fields. We now apply our guiding and gyrocentre transformations 
to write this many-body Lagrangian in terms of the particle gyrocentre variables. The system Lagrangian, which is the sum of the particle Lagrangians,
plus the field component integrated over space, then directly leads to gyrocentre Euler-Lagrange equations, and Poisson and Amp\`{e}re equations for the fields. 
We are usually interested in the smooth limit of these equations (potentially with a collision operator representing short spatial scale correlations), with particles described 
by a distribution function $\bar{F}(\bar{Z})$, in which case the time evolution of $\bar{F}$ can be evaluated in terms of the Euler-Lagrange equations
of the gyroparticles (a gyrokinetic Vlasov equation) and in field equations sums over particles are replaced by integrals of $\bar{F}$.

We note the contrast between this approach, which is similar to that of Refs. \onlinecite{Miyato2009} and \onlinecite{Scott2010}, and attempts to vary a system Lagrangian written in terms of the distribution function: the Euler-Lagrange equations appear naturally, rather than being inserted by hand as a constraint.

At this point it is useful to introduce some notation: we denote a mapping from coordinate system $\bar{Z}$ to $z$ as $\mathcal{T}_{\bar{Z}\to z}$ and the associated Jacobian as $J_{\bar{Z}\to z} = |\bar{\partial}_i\mathcal{T}_{\bar{Z}\to z}\bar{Z}_j|$.

We will consider only the electrostatic, quasineutral limit where the field terms have been ignored and species sums, charges and masses have been suppressed. The Poisson equation can be written as the variation of the system Lagrangian in original coordinates with respect to $\phi$, and this can also be written directly in gyrocentre coordinates, based on the above consideration of interpretation as 
the limit of a many body theory,
\begin{equation}
\frac{\partial}{\partial\phi}\int{\rm d}^6zf(z)L_{\rm p}(z)=\frac{\partial}{\partial\phi}\int{\rm d}^6\bar{Z}\bar{F}(\bar{Z})L_{\rm p}(\bar{Z});
\label{l}
\end{equation}
the invariance of the value is also what we expect due to the covariance of the form of the integral. Note, however, that, here, $f$ must be defined so that it transforms as a scalar density: the `usual' gyrocentre distribution function is actually $\bar{F}'(\bar{Z}) = f(\mathcal{T}_{\bar{Z} \to z} \bar{Z} ) = (J_{\bar{Z} \to z})^{-1}  \bar{F}(\bar{Z})$. This Jacobian is a function of $\phi$, unlike for the transformations in the weak-flow case, and varying $\phi$ with fixed $\bar{F}$ is 
not identical to varying $\phi$ with fixed $\bar{F}'$.

Performing this variation yields
\begin{equation}
\begin{aligned}
0 = (\delta L)_\phi=&-\int{\rm d}^3r\delta\phi(\bm{r})\int{\rm d}^6\bar{Z}\delta(\bar{\bm{R}}+\bar{\bm{\rho}}-\bm{r})[(1\\&
+\Omega^{-2}\bar{\bm{\nabla}}\tilde{\Phi}\times\hat{\bm{b}}\cdot\bar{\bm{\nabla}}+\Omega^{-1}\tilde{\phi}\partial_{\bar{\mu}})\bar{F}\\&
+\Omega^{-1}\hat{\bm{b}}\cdot\bar{\bm{\nabla}}\times(\bar{F}\dot{\bar{\bm{R}}}-2\bar{F}\bar{\bm{u}})].
\end{aligned}
\label{dl}
\end{equation}
If the distribution function $\bar{F}'$ is uniform, and we neglect terms which are of order $\epsilon^2$, this Poisson equation reduces to the usual weak-flow Poisson equation as shown in Appendix \ref{appendix_poisson}. 

For weak flows, it has been shown\cite{Brizard2007} that the variational method for obtaining the Poisson equation is equivalent to the direct method of
setting the charge-density to zero, up to the chosen order of approximation. Here, we have the quasineutrality equation
\begin{equation}
0=\int{\rm d}^6z\delta(\bm{x}-\bm{r})f(z),\label{nd}
\end{equation}
where $f$ is the original distribution function. A change of variables can be made to guiding-centre coordinates, and the guiding-centre
distribution function $F'(Z)$ can be expressed in terms of the gyrocentre distribution function $\bar{F}'(Z)$ using the Lie transform,\cite{Dubin1983} to yield
\begin{equation}
0=\int J_{Z\to z}{\rm d}^6Z\delta(\bm{R}+\bm{\rho}-\bm{r})\mathsf{T}\bar{F}'.\label{nd1}
\end{equation}
Note that the Jacobian is of the transform from original coordinates to guiding-centre space, which is not equal to $J_{\bar{Z} \to z}$ for this strong-flow formalism; the two are equivalent in the weak-flow analysis\footnote{The Jacobians, which can be written as the square root of the determinant of the appropriate Lagrange matrix, are only a function of the symplectic part of the Lagragian, which is unperturbed and unmodified by the Lie transform for the weak- but not the strong-flow formalism.}. Explicit evaluation of Eq. \eqref{nd1} leads to the same result as the variational formalism; details are given in Appendix \ref{appendix_poisson}
for completeness.

Alternatively, we can directly evaluate Eq. \eqref{nd} in gyrocentre coordinates so that the Lie transform appears in the delta function: this
again gives an equivalent expression for the Poisson equation. 
\section{Numerical solution of the equations}
The second order Lagrangian derived here allows relatively simple explicit forms of the equations of motion for the particles, and the Poisson equation is also of a tractable form. However, the advection of gyroscale structures with velocities of order $v_{\rm t}$ results in time variations of order of the gyration time, and standard Eulerian schemes would be forced to run on this time scale. This would negate the point of using gyrokinetics, and appears suboptimal considering that nonlinear time scales are expected to be of the order of the inverse vorticity. This suggests the use of semi-Lagrangian or particle-in-cell (PIC) methods which allow Courant numbers much larger than one. We have chosen to use a PIC method for the particle distribution and a finite-difference method for the field equations.

The dependence of the Euler-Lagrange equations derived from the first or second order Lagrangian on the time derivative of the potential implies that the Euler-Lagrange equations and the gyrokinetic Poisson equation must be solved simultaneously in general: this complication arises because part of the polarisation drift is now contained within the particle trajectories, unlike in the weak-flow gyrokinetic formalism where the polarisation drift is captured completely in the change of variables. The Poisson equation is also a first order differential equation for the potential: however, the term containing the derivative of the potential is of a smaller order than the dominant terms, and we solve this equation in the quasi-static limit (the solution is the smooth continuation of the solution in the limit $\epsilon \rightarrow 0$).    

Our current approach to solving these equations is to expand the Poisson equation around an approximate solution $\bar{F}_0'$. The polarisation of the background part of the plasma $\bar{F}_0'$ is directly captured in the Poisson equation, and the $\delta\bar{F}'$ contribution is comparatively small, and can be added as a correction. The Euler-Lagrange and Poisson equations are then solved iteratively, with the first particle trajectory step neglecting the polarisation term, given that only the electrostatic potential, and not its derivative, are known at this point. Iteration is continued to include both the effects of background and $\delta\bar{F}'$ polarisation; the convergence ratio per iteration is of order $\epsilon$. More general choices for $\bar{F}'_0$ are possible which allow large spatio-temporal fluctuations in particle densities and the algorithm appears to be generalisable.
\appendix\section{Guiding-centre Lagrangian}\label{agc}
Using Eq. \eqref{gt},
\begin{align*}
\gamma=&[\bm{A}(\bm{R}+\bm{\rho})+U\hat{\bm{b}}+\bm{v}_\perp+\bm{u}]\cdot{\rm d}(\bm{R}+\bm{\rho})\\&
-\big[\tfrac{1}{2}(U\hat{\bm{b}}+\bm{v}_\perp+\bm{u})^2+\phi\big]{\rm d}t,
\end{align*}
which can be expanded as
\begin{align*}
\gamma=&[\bm{A}(\bm{R})+(\bm{\rho}\cdot\bm{\nabla})\bm{A}(\bm{R})+U\hat{\bm{b}}+\bm{v}_\perp+\bm{u}]\cdot({\rm d}\bm{R}+{\rm d}\bm{\rho})\\&
-\big[\tfrac{1}{2}(\bm{u}+\bm{v}_\perp+U\hat{\bm{b}})^2+\phi\big]{\rm d}t
\end{align*}
and rearranged as
\begin{align*}
\gamma=&[\bm{A}(\bm{R})+U\hat{\bm{b}}+\bm{u}]\cdot{\rm d}\bm{R}+\bm{A}(\bm{R})\cdot{\rm d}\bm{\rho}+[(\bm{\rho}\cdot\bm{\nabla})\bm{A}(\bm{R})\\&
+\bm{v}_\perp]\cdot{\rm d}\bm{R}+\bm{u}\cdot{\rm d}\bm{\rho}+[(\bm{\rho}\cdot\bm{\nabla})\bm{A}(\bm{R})+\bm{v}_\perp]\cdot{\rm d}\bm{\rho}\\&
-\big(\tfrac{1}{2}U^2+\mu\Omega+\tfrac{1}{2}\bm{u}^2+\langle\phi\rangle+\delta_1\tilde{\phi}\big){\rm d}t\\&
+{\rm d}\left\{-\bm{\rho}\cdot\left[\big(\tfrac{1}{2}\bm{\rho}\cdot\bm{\nabla}+1\big)\bm{A}(\bm{R})+\bm{u}\right]\right\},
\end{align*}
where we have used
\[\hat{\bm{b}}\cdot\bm{u}=0.\]
Using integration by parts,
\begin{align*}
\gamma=&\gamma_{\bm{R}}\cdot{\rm d}\bm{R}-{\rm d}\bm{A}(\bm{R})\cdot\bm{\rho}+[(\bm{\rho}\cdot\bm{\nabla})\bm{A}(\bm{R})+\bm{\rho}\times\bm{\Omega}]\cdot{\rm d}\bm{R}\\&
-\bm{\rho}\cdot{\rm d}\bm{u}+[\rho(\hat{\bm{\rho}}\cdot\bm{\nabla})\bm{A}(\bm{R})+\bm{v}_\perp]\cdot(\bm{\rho}_{,v_\perp}{\rm d}v_\perp+\bm{\rho}_{,\theta}{\rm d}\theta)\\&
-v_\perp\Omega^{-2}(\hat{\bm{\rho}}\cdot\bm{\nabla})\bm{A}(\bm{R})\cdot\hat{\bm{\rho}}{\rm d}v_\perp\\&
-\mu\Omega^{-1}[(\hat{\bm{\rho}}\cdot\bm{\nabla})\bm{A}(\bm{R})\cdot\hat{\bm{\rho}}]_{,\theta}{\rm d}\theta+\gamma_t{\rm d}t,
\end{align*}
\[\gamma_{\bm{R}}=\bm{A}(\bm{R})+U\hat{\bm{b}}+\bm{u},\]
\[\gamma_t=-\tfrac{1}{2}U^2-\mu\Omega-\tfrac{1}{2}\bm{u}^2-\langle\phi\rangle-\delta_1\tilde{\phi}.\]
Rewriting,
\begin{align*}
\gamma=&\gamma_{\bm{R}}\cdot{\rm d}\bm{R}-({\rm d}\bm{R}\cdot\bm{\nabla})\bm{A}(\bm{R})\cdot\bm{\rho}+\{(\bm{\rho}\cdot\bm{\nabla})\bm{A}(\bm{R})+\\&
\bm{\rho}\times[\bm{\nabla}\times\bm{A}(\bm{R})]\}\cdot{\rm d}\bm{R}-\bm{\rho}\cdot{\rm d}\bm{u}\\&
+[v_\perp\Omega^{-1}(\hat{\bm{\rho}}\cdot\bm{\nabla})\bm{A}(\bm{R})+\bm{v}_\perp]\cdot(\hat{\bm{\rho}}\rho_{,v_\perp}{\rm d}v_\perp+\rho\hat{\bm{\rho}}_{,\theta}{\rm d}\theta)\\&
-v_\perp\Omega^{-2}(\hat{\bm{\rho}}\cdot\bm{\nabla})\bm{A}(\bm{R})\cdot\hat{\bm{\rho}}{\rm d}v_\perp\\&
-\mu\Omega^{-1}[(\hat{\bm{v}}_\perp\cdot\bm{\nabla})\bm{A}(\bm{R})\cdot\hat{\bm{\rho}}+(\hat{\bm{\rho}}\cdot\bm{\nabla})\bm{A}(\bm{R})\cdot\hat{\bm{v}}_\perp]{\rm d}\theta\\&
+\gamma_t{\rm d}t.
\end{align*}
Using the vector identity
\begin{align*}
\bm{\nabla}[\bm{A}(\bm{R})\cdot\bm{\rho}]=&[\bm{A}(\bm{R})\cdot\bm{\nabla}]\bm{\rho}+(\bm{\rho}\cdot\bm{\nabla})\bm{A}(\bm{R})\\&
+\bm{A}(\bm{R})\times(\bm{\nabla}\times\bm{\rho})+\bm{\rho}\times[\bm{\nabla}\times\bm{A}(\bm{R})],
\end{align*}
\begin{align*}
\gamma=&\gamma_{\bm{R}}\cdot{\rm d}\bm{R}-\bm{\rho}\cdot{\rm d}\bm{u}+\{2-\Omega^{-1}[(\hat{\bm{v}}_\perp\cdot\bm{\nabla})\bm{A}(\bm{R})\cdot\hat{\bm{\rho}}\\&
-(\hat{\bm{\rho}}\cdot\bm{\nabla})\bm{A}(\bm{R})\cdot\hat{\bm{v}}_\perp]\}\mu{\rm d}\theta+\gamma_t{\rm d}t
\end{align*}
and, by evaluating the $\theta$-component in the same way as is done the weak-flow gyrokinetic formalism, we obtain Eq. \eqref{gc}.
\section{Gyrocentre Lagrangian}\label{agy}
The requirement
\[\delta_1\tilde{\phi}=O(\epsilon)\]
is equivalent to restrictions on the choices of $\theta$-independent potential appearing in Eq. \eqref{gc} and $\bm{u}$ given by\begin{equation}
\phi_{\rm g}-\phi(\bm{R})\leq O(\epsilon)\label{op}\end{equation}
and\begin{equation}
\bm{u}-\Omega^{-1}\hat{\bm{b}}\times\bm{\nabla}\phi(\bm{R})\leq O(\epsilon),\label{ou}\end{equation}
respectively, where $\phi_{\rm g}$ is a general $\theta$-independent potential. Some choices of $\phi_{\rm g}$ and $\bm{u}$ that satisfy orderings \eqref{op} and \eqref{ou} are $\phi_{\rm g}=\phi(\bm{R})$, $\phi_{\rm g}=\langle\phi\rangle$,
\[\bm{u}=\frac{1}{\Omega}\hat{\bm{b}}\times\bm{\nabla}\phi(\bm{R})\]
and\[
\bm{u}=\frac{1}{\Omega}\hat{\bm{b}}\times\bm{\nabla}\langle\phi\rangle.\]
Using Eq. \eqref{lm}, the non-zero zeroth-order Lagrange matrix components are
\begin{align*}
&\omega_{0R_{i'}R_{j'}}=\epsilon_{i'j'k'}\Omega^*_{k'},\\
&\omega_{0\bm{R}\mu}=-\frac{\partial\bm{u}}{\partial\mu},\\
&\omega_{0\bm{R}t}=-\bm{\nabla}\langle\phi\rangle-\bm{u}\times(\bm{\nabla}\times\bm{u})-\left(\bm{u}\cdot\bm{\nabla}+\frac{\partial}{\partial t}\right)\bm{u},\\
&\omega_{0\mu t}=-\frac{\partial\langle\phi\rangle}{\partial\mu}-\bm{u}\cdot\frac{\partial\bm{u}}{\partial\mu}-\Omega,\\
&\omega_{0\bm{R}U}=-\hat{\bm{b}},\\
&\omega_{0Ut}=-U,\\
&\omega_{0\mu\theta}=1,
\end{align*}
where $i',j',k'\in\{1,2,3\}$ and we have used $\bm{u}=\bm{u}(\bm{R},\mu,t)$. The first-order part of the gyrocentre Lagrangian is
\begin{eqnarray*}
\bar{\Gamma}_1&=&\Gamma_1-L_1\Gamma_0+{\rm d}S_1\\&
=&-\bm{\rho}\cdot\left({\rm d}\bm{R}\cdot\bm{\nabla}+{\rm d}\mu\frac{\partial}{\partial\mu}+{\rm d}t\frac{\partial}{\partial t}\right)\bm{u}-\delta_1\tilde{\phi}{\rm d}t\\&&
+g^{\bm{R}}_1\cdot\left\{\bm{\Omega}^*\times{\rm d}\bm{R}+\frac{\partial\bm{u}}{\partial\mu}{\rm d}\mu+\left[\bm{\nabla}\langle\phi\rangle+\bm{u}\times(\bm{\nabla}\times\bm{u})\right.\right.\\&&\left.\left.
+\left(\bm{u}\cdot\bm{\nabla}+\frac{\partial}{\partial t}\right)\bm{u}\right]{\rm d}t\right\}\\&&
-g^\mu_1\left[\frac{\partial\bm{u}}{\partial\mu}\cdot{\rm d}\bm{R}+{\rm d}\theta-\left(\frac{\partial\langle\phi\rangle}{\partial\mu}+\bm{u}\cdot\frac{\partial\bm{u}}{\partial\mu}+\Omega\right){\rm d}t\right]\\&&
+g^{\theta}_1{\rm d}\mu+{\rm d}S_1.
\end{eqnarray*}
Refactorising,
\begin{align*}
\bar{\Gamma}_1=&{\rm d}\bm{R}\cdot\left(-\bm{\rho}\cdot\bm{\nabla}\bm{u}+g^{\bm{R}}_1\times\bm{\Omega}^*-g^\mu_1\frac{\partial\bm{u}}{\partial\mu}+\bm{\nabla}S_1\right)\\&
+{\rm d}\mu\left(-\bm{\rho}\cdot\frac{\partial\bm{u}}{\partial\mu}+g^{\bm{R}}_1\cdot\frac{\partial\bm{u}}{\partial\mu}+g^\theta_1+\frac{\partial S_1}{\partial\mu}\right)\\&
+{\rm d}\theta\left(-g^\mu_1+\frac{\partial S_1}{\partial\theta}\right)+{\rm d}t\left\{-\bm{\rho}\cdot\frac{\partial\bm{u}}{\partial t}-\delta_1\tilde{\phi}\right.\\&+g^{\bm{R}}_1\cdot\left[\bm{\nabla}\langle\phi\rangle
+\bm{u}\times(\bm{\nabla}\times\bm{u})+\left(\bm{u}\cdot\bm{\nabla}+\frac{\partial}{\partial t}\right)\bm{u}\right]\\&\left.
+g^\mu_1\left(\frac{\partial\langle\phi\rangle}{\partial\mu}+\bm{u}\cdot\frac{\partial\bm{u}}{\partial\mu}+\Omega\right)+\frac{\partial S_1}{\partial t}\right\}.
\end{align*}
Solving for \(g_1\) in terms of \(S_1\) such that \(\bar{\Gamma}_1\) is only composed of a first-order time component,
\begin{align*}
\bar{\Gamma}_1=&\left\{-\bm{\rho}\cdot\frac{\partial\bm{u}}{\partial t}-\delta_1\tilde{\phi}+\frac{1}{\Omega}\left[\bm{\rho}\cdot\left(\hat{\bm{b}}\times\bm{\nabla}\right)\bm{u}\right.\right.\\&\left.\left.
+\bm{\nabla}S_1\times\hat{\bm{b}}\right]\cdot\bm{\nabla}\langle\phi\rangle+\Omega\frac{\partial S_1}{\partial\theta}+\frac{\partial S_1}{\partial t}\right\}{\rm d}t+O\left(\epsilon^2\right),
\end{align*}
yields the non-zero \(g_1\) components
\begin{align*}
&g^{\bm{R}}_1=\frac{1}{\Omega}\left[\bm{\rho}\cdot\left(\hat{\bm{b}}\times\bm{\nabla}\right)\bm{u}+\bm{\nabla}S_1\times\hat{\bm{b}}\right],\\
&g^\mu_1=\frac{\partial S_1}{\partial\theta},\\
&g^\theta_1=\bm{\rho}\cdot\frac{\partial\bm{u}}{\partial\mu}-\frac{\partial S_1}{\partial\mu}.\\
\end{align*}
Rearranging,
\begin{align*}
\bar{\Gamma}_1=&\left[-\bm{\rho}\cdot\left(\frac{\partial}{\partial t}+\frac{1}{\Omega}\hat{\bm{b}}\times\bm{\nabla}\langle\phi\rangle\cdot\bm{\nabla}\right)\bm{u}-\delta_1\tilde{\phi}\right.\\&\left.
+\left(\frac{1}{\Omega}\hat{\bm{b}}\times\bm{\nabla}\langle\phi\rangle\cdot\bm{\nabla}+\frac{\partial}{\partial t}\right)S_1+\Omega\frac{\partial S_1}{\partial\theta}\right]{\rm d}t+O\left(\epsilon^2\right)\\
=&\left(-\delta_1\tilde{\phi}+\Omega\frac{\partial S_1}{\partial\theta}\right){\rm d}t+O\left(\epsilon^2\right),
\end{align*}
where the last equality follows from Ref. \onlinecite{Dimits2010} and gives
\[S_1=\frac{\delta_1\tilde{\Phi}}{\Omega}.\]
$\Gamma_1$ yields
\begin{align*}&
\omega_{1\bm{R}\mu}=\bm{\nabla}\bm{u}\cdot\partial_\mu\bm{\rho},\\&
\omega_{1\bm{R}\theta}=\partial_\theta\bm{\rho}\cdot\bm{\nabla}\bm{u},\\&
\omega_{1\bm{R}t}=-\bm{\nabla}\delta_1\tilde{\phi},\\&
\omega_{1\mu\theta}=\partial_\theta\bm{\rho}\cdot\partial_\mu\bm{u},\\&
\omega_{1\mu t}=-\partial_t\bm{u}\cdot\partial_\mu\bm{\rho}-\partial_\mu\delta_1\tilde{\phi},\\&
\omega_{1\theta t}=-\partial_\theta\left(\bm{\rho}\cdot\partial_t\bm{u}+\delta_1\tilde{\phi}\right)
\end{align*}
and the expression for $\bar{\Gamma}_2$\cite{Littlejohn1982} is
\begin{align*}
\bar{\Gamma}_2&=\Gamma_2-L_1\Gamma_1+\left(\tfrac{1}{2}L^2_1-L_2\right)\Gamma_0+{\rm d}S_2\\&
=\Gamma_2-L_1\Gamma_1+\tfrac{1}{2}L_1(L_1\Gamma_0)-L_2\Gamma_0+{\rm d}S_2\\&
=\Gamma_2-L_1\Gamma_1+\tfrac{1}{2}L_1(\Gamma_1+{\rm d}S_1-\bar{\Gamma}_1)-L_2\Gamma_0+{\rm d}S_2\\&
=\Gamma_2-\tfrac{1}{2}L_1\Gamma_1-L_2\Gamma_0+{\rm d}S_2,
\end{align*}
where $L_1{\rm d}S_1=\bar{\Gamma}_1=0$,
\begin{align*}
\Gamma_2=&\left[g^{\bm{R}}_1\times(\bm{\nabla}\times\bm{u})-g^\mu_1\partial_\mu\bm{u}\right]\cdot{\rm d}\bm{R}-g^{\bm{R}}_1\partial_\mu\bm{u}{\rm d}\mu\\&
+\left\{g^{\bm{R}}_1\cdot\bm{u}\times(\bm{\nabla}\times\bm{u})+g^\mu_1\left(\partial_\mu\langle\phi\rangle+\bm{u}\cdot\partial_\mu\bm{u}\right)\right.\\&\left.
+(\partial_t+\bm{u}\cdot\bm{\nabla})(S_1-\bm{\rho}\cdot\bm{u})\right\}{\rm d}t+O\left(\epsilon^3\right),
\end{align*}
\begin{align*}
-\tfrac{1}{2}L_1\Gamma_1=&\tfrac{1}{2}\left\{\left(g^\mu_1\bm{\nabla}\bm{u}\cdot\partial_\mu\bm{\rho}+g^\theta_1\partial_\theta\bm{\rho}\cdot\bm{\nabla}\bm{u}\right)\cdot{\rm d}\bm{R}\vphantom{\tilde{\phi}}\right.\\&\left.
+\left(g^\theta_1\partial_\theta\bm{\rho}\cdot\partial_\mu\bm{u}-g^{\bm{R}}_1\cdot\bm{\nabla}\bm{u}\cdot\partial_\mu\bm{\rho}\right){\rm d}\mu\right.\\&\left.
-\left(g^{\bm{R}}_1\cdot\bm{\nabla}\bm{u}\cdot\partial_\theta\bm{\rho}+g^\mu_1\partial_\theta\bm{\rho}\cdot\partial_\mu\bm{u}\right){\rm d}\theta\right.\\&\left.
+\left[g^{\bm{R}}_1\cdot\bm{\nabla}\delta_1\tilde{\phi}+g^\mu_1\left(\partial_t\bm{u}\cdot\partial_\mu\bm{\rho}+\partial_\mu\delta_1\tilde{\phi}\right)\right.\right.\\&\left.\left.
+g^\theta_1\left(\partial_\theta\bm{\rho}\cdot\partial_t\bm{u}+\partial_\theta\delta_1\tilde{\phi}\right)\right]{\rm d}t\right\},
\end{align*}
\begin{align*}
-L_2\Gamma_0=&g^{\bm{R}}_2\times\bm{\Omega}\cdot{\rm d}\bm{R}+g^\theta_2{\rm d}\mu
-g^\mu_2{\rm d}\theta+\left(g^{\bm{R}}_2\cdot\bm{\nabla}\langle\phi\rangle\right.\\&\left.
+g^\mu_2\Omega\right){\rm d}t+O\left(\epsilon^3\right)
\end{align*}
and
\[{\rm d}S_2=\bm{\nabla}S_2\cdot{\rm d}\bm{R}+\partial_US_2{\rm d}U+\partial_\mu S_2{\rm d}\mu+\partial_\theta S_2{\rm d}\theta+\partial_tS_2{\rm d}t.\]
In order to facilitate several cancelations during the computation of the second-order gyrocentre Lagrangian, we choose $\bm{u}$ to be the $\rm E\times B$ drift velocity associated with the $\theta$-independent potential that appears in Eq. \eqref{gc}. Solving for \(g_2\) in terms of \(S_2\) such that \(\bar{\Gamma}_2\) is only composed of a time component,
\begin{align*}
\bar{\Gamma}_2=&\left[g^\mu_1\partial_\mu\langle\phi\rangle+(\partial_t+\bm{u}\cdot\bm{\nabla})(S_1-\bm{\rho}\cdot\bm{u})+\tfrac{1}{2}g^a_1\left(\partial_a\delta_1\tilde{\phi}\right.\right.\\&\left.\left.
-\Omega\partial_\theta\bm{\rho}\cdot\partial_a\bm{u}\right)+\Omega\partial_\theta S_2\vphantom{\tilde{\phi}}\right]{\rm d}t+O\left(\epsilon^3\right),
\end{align*}
yields the non-zero \(g_2\) components
\begin{align*}
g^{\bm{R}}_2=&\Omega^{-1}\left[g^{\bm{R}}_1\times(\bm{\nabla}\times\bm{u})-g^\mu_1\partial_\mu\bm{u}+\tfrac{1}{2}g^a_1\bm{\nabla}\bm{u}\cdot\partial_a\bm{\rho}\right.\\&\left.
+\bm{\nabla}S_2\vphantom{\tfrac{1}{2}}\right]\times\hat{\bm{b}},
\end{align*}
\[g^\mu_2=\partial_\theta S_2-\tfrac{1}{2}g^a_1\partial_\theta\bm{\rho}\cdot\partial_a\bm{u},\]
\[g^\theta_2=g^{\bm{R}}_1\partial_\mu\bm{u}-\tfrac{1}{2}\left(g^\theta_1\partial_\theta\bm{\rho}\cdot\partial_\mu\bm{u}-g^{\bm{R}}_1\cdot\bm{\nabla}\bm{u}\cdot\partial_\mu\bm{\rho}\right)-\partial_\mu S_2.\]
Using the freedom of $S_2$ to remove the $\theta$-dependent terms in $\bar{\Gamma}_2$,
\begin{eqnarray}
\bar{\Gamma}_2&=&\tfrac{1}{2}\left\langle g^a_1\left(\partial_a\delta_1\tilde{\phi}-\Omega\partial_\theta\bm{\rho}\cdot\partial_a\bm{u}\right)\right\rangle{\rm d}t\nonumber\\&
=&\tfrac{1}{2}\left\langle g^a_1\left(\partial_a\tilde{\phi}+\Omega\bm{u}\cdot\partial_a\partial_\theta\bm{\rho}\right)\right\rangle{\rm d}t\nonumber\\&
=&\big[\tfrac{1}{2}\langle g^{\bm{R}}_1\cdot\bm{\nabla}\tilde{\phi}\rangle+\tfrac{1}{2}\Omega^{-1}\langle\delta_1\tilde{\phi}^2\rangle_{,\mu}+\hat{\bm{b}}\times\langle\delta_1\tilde{\phi}\bm{\rho}\rangle\cdot\bm{u}_{,\mu}\big]{\rm d}t\nonumber\\&
=&\big[\tfrac{1}{2}\langle g^{\bm{R}}_1\cdot\bm{\nabla}\tilde{\phi}\rangle+\tfrac{1}{2}\Omega^{-1}\langle\tilde{\phi}^2\rangle_{,\mu}-\bm{u}\cdot\hat{\bm{b}}\times\langle\tilde{\phi}\bm{\rho}\rangle_{,\mu}\nonumber\\&&
+\tfrac{1}{2}\bm{u}^2\big]{\rm d}t\label{g2}\\&
=&\big[\tfrac{1}{2}\langle g_1^{\bm{R}}\cdot\bm{\nabla}\tilde{\phi}\rangle+\tfrac{1}{2}\Omega^{-1}\partial_\mu\langle\tilde{\phi}^2\rangle-\tfrac{1}{2}\bm{u}^2\big]{\rm d}t.\nonumber\end{eqnarray}
\section{Euler-Lagrange equations}\label{ael}
The Euler-Lagrange equations in terms of gyrocentre coordinates,
\[\bar{\omega}_{ij}\dot{\bar{Z}}_j=\bar{\omega}_{ti},\]
give
\[\bar{\omega}_{\bar{U}\bar{\bm{R}}}\dot{\bar{\bm{R}}}=\bar{\omega}_{t\bar{U}}\]
for \(i=\bar{U}\). Substituting,
\[\hat{\bm{b}}\cdot\dot{\bar{\bm{R}}}=\bar{U}.\]
For \(i=\bar{R}_{i'}\),
\[\bar{\omega}_{\bar{R}_{i'}\bar{R}_{j'}}\dot{\bar{R}}_{j'}+\bar{\omega}_{\bar{R}_{i'}\bar{U}}\dot{\bar{U}}=\bar{\omega}_{t\bar{R}_{i'}}.\]
Substituting,
\[\dot{\bar{\bm{R}}}\times\bar{\bm{\Omega}}^*-\dot{\bar{U}}\hat{\bm{b}}=\bar{\omega}_{t\bar{\bm{R}}}.\]
Taking the cross product with \(\hat{\bm{b}}\),
\[\hat{\bm{b}}\times\left(\dot{\bar{\bm{R}}}\times\bar{\bm{\Omega}}^*-\dot{\bar{U}}\hat{\bm{b}}\right)=\hat{\bm{b}}\times\bar{\omega}_{t\bar{\bm{R}}}.\]
Using the vector identity
\begin{align*}
&\hat{\bm{b}}\times\left(\dot{\bar{\bm{R}}}\times\bar{\bm{\Omega}}^*\right)=\left(\hat{\bm{b}}\cdot\bar{\bm{\Omega}}^*\right)\dot{\bar{\bm{R}}}-\left(\hat{\bm{b}}\cdot\dot{\bar{\bm{R}}}\right)\bar{\bm{\Omega}}^*,\\
&\dot{\bar{\bm{R}}}=\frac{1}{\bar{\Omega}^*_\|}\left(\hat{\bm{b}}\times\bar{\omega}_{t\bar{\bm{R}}}+\bar{U}\bar{\bm{\Omega}}^*\right).
\end{align*}
Upon substituting and using that
\[\hat{\bm{b}}\times\left[\hat{\bm{b}}\times(\bar{\bm{\nabla}}\times\bar{\bm{u}})\right]=\left[\hat{\bm{b}}\cdot(\bar{\bm{\nabla}}\times\bar{\bm{u}})\right]\hat{\bm{b}}-\left(\hat{\bm{b}}\cdot\hat{\bm{b}}\right)\bar{\bm{\nabla}}\times\bar{\bm{u}},\]
\begin{align*}
\dot{\bar{\bm{R}}}=&\frac{1}{\bar{\Omega}^*_\|}\left(\hat{\bm{b}}\times\left[\bar{\bm{\nabla}}\langle\phi\rangle+\bar{\bm{u}}\times(\bar{\bm{\nabla}}\times\bar{\bm{u}})+\left(\bar{\bm{u}}\cdot\bar{\bm{\nabla}}+\frac{\partial}{\partial t}\right)\bar{\bm{u}}\right]\right.\\&\left.
+\bar{U}\left\{\bm{\Omega}+\left[\hat{\bm{b}}\cdot\bar{\bm{\nabla}}\times\bar{\bm{u}}\right]\hat{\bm{b}}-\hat{\bm{b}}\times\left[\hat{\bm{b}}\times(\bar{\bm{\nabla}}\times\bar{\bm{u}})\right]\right\}\vphantom{\frac{\partial}{\partial t}}\right).
\end{align*}
Using that
\begin{equation}
\hat{\bm{b}}\times[\bar{\bm{u}}\times(\bar{\bm{\nabla}}\times\bar{\bm{u}})]=\left[\hat{\bm{b}}\cdot(\bar{\bm{\nabla}}\times\bar{\bm{u}})\right]\bar{\bm{u}}-\left(\hat{\bm{b}}\cdot\bar{\bm{u}}\right)\bar{\bm{\nabla}}\times\bar{\bm{u}},\label{i}
\end{equation}
and that
\begin{align*}
\bar{\bm{\nabla}}\left(\hat{\bm{b}}\cdot\bar{\bm{u}}\right)=&\left(\hat{\bm{b}}\cdot\bar{\bm{\nabla}}\right)\bar{\bm{u}}+(\bar{\bm{u}}\cdot\bar{\bm{\nabla}})\hat{\bm{b}}+\hat{\bm{b}}\times(\bar{\bm{\nabla}}\times\bar{\bm{u}})\\&
+\bar{\bm{u}}\times\left(\bar{\bm{\nabla}}\times\hat{\bm{b}}\right),
\end{align*}
\begin{align*}
\dot{\bar{\bm{R}}}=&\bar{\bm{u}}+\frac{1}{\bar{\Omega}^*_\|}\hat{\bm{b}}\times\left(\frac{\partial}{\partial t}+\bar{\bm{u}}\cdot\bar{\bm{\nabla}}+\bar{U}\bar{\nabla}_\|\right)\bar{\bm{u}}+\bar{U}\hat{\bm{b}}\\&
+\frac{1}{\bar{\Omega}^*_\|}\left(\hat{\bm{b}}\times\bar{\bm{\nabla}}\langle\phi\rangle-\Omega\bar{\bm{u}}\right).
\end{align*}
Instead of taking the cross product with \(\hat{\bm{b}}\), projecting onto \(\bar{\bm{\Omega}}^*\),
\[\bar{\bm{\Omega}}^*\cdot\left(\dot{\bar{\bm{R}}}\times\bar{\bm{\Omega}}^*-\dot{\bar{U}}\hat{\bm{b}}\right)=\bar{\bm{\Omega}}^*\cdot\omega_{t\bar{\bm{R}}}\]
and
\begin{align*}
\dot{\bar{U}}=&\frac{1}{\bar{\Omega}^*_\|}\bar{\bm{\Omega}}^*\cdot\bar{\omega}_{\bar{\bm{R}}t}\\
=&-\frac{1}{\bar{\Omega}^*_\|}\left(\bar{\Omega}^*_\|\hat{\bm{b}}+\hat{\bm{b}}\times\bar{\nabla}_\|\bar{\bm{u}}\right)\cdot\left[\bar{\bm{\nabla}}\langle\phi\rangle+\bar{\bm{u}}\times(\bar{\bm{\nabla}}\times\bar{\bm{u}})\right.\\&\left.
+\left(\bar{\bm{u}}\cdot\bar{\bm{\nabla}}+\frac{\partial}{\partial t}\right)\bar{\bm{u}}\right].
\end{align*}
Using that
\[\tfrac{1}{2}\bar{\bm{\nabla}}(\bar{\bm{u}}\cdot\bar{\bm{u}})=(\bar{\bm{u}}\cdot\bar{\bm{\nabla}})\bar{\bm{u}}+\bar{\bm{u}}\times(\bar{\bm{\nabla}}\times\bar{\bm{u}})\]
and Equation (\ref{i}),
\begin{align*}
\dot{\bar{U}}=&-\bar{\nabla}_\|\langle\phi\rangle+\frac{1}{\bar{\Omega}^*_\|}\left(\bar{\nabla}_\|\bar{\bm{u}}\right)\cdot\hat{\bm{b}}\times\left(\frac{\partial}{\partial t}+\bar{\bm{u}}\cdot\bar{\bm{\nabla}}\right)\bar{\bm{u}}\\&
+\frac{1}{\bar{\Omega}^*_\|}\left(\hat{\bm{b}}\times\bar{\bm{\nabla}}\langle\phi\rangle-\Omega\bar{\bm{u}}\right)\cdot\bar{\nabla}_\|\bar{\bm{u}}\\&
+\frac{\Omega-\bar{\Omega}^*_\|+\hat{\bm{b}}\cdot\bar{\bm{\nabla}}\times\bar{\bm{u}}}{\bar{\Omega}^*_\|}\bar{\bm{u}}\cdot\bar{\nabla}_\|\bar{\bm{u}}\\
=&-\bar{\nabla}_\|\langle\phi\rangle+\frac{1}{\bar{\Omega}^*_\|}\left(\bar{\nabla}_\|\bar{\bm{u}}\right)\cdot\hat{\bm{b}}\times\left(\frac{\partial}{\partial t}+\bar{\bm{u}}\cdot\bar{\bm{\nabla}}\right)\bar{\bm{u}}\\&
+\frac{1}{\bar{\Omega}^*_\|}\left(\hat{\bm{b}}\times\bar{\bm{\nabla}}\langle\phi\rangle-\Omega\bar{\bm{u}}\right)\cdot\bar{\nabla}_\|\bar{\bm{u}}.
\end{align*}
For \(i=\bar{\theta}\),
\[\bar{\omega}_{\bar{\theta}\bar{\mu}}\dot{\bar{\mu}}=\bar{\omega}_{t\bar{\theta}}.\]
Substituting,
\[\dot{\bar{\mu}}=0.\]
For \(i=\bar{\mu}\),
\[\bar{\omega}_{\bar{\mu}\bar{\bm{R}}}\dot{\bar{\bm{R}}}+\bar{\omega}_{\bar{\mu}\bar{\theta}}\dot{\bar{\theta}}=\bar{\omega}_{t\bar{\mu}}.\]
Substituting,
\begin{align*}
\dot{\bar{\theta}}=&\Omega+\frac{\partial\langle\phi\rangle}{\partial\bar{\mu}}-\frac{1}{\bar{\Omega}^*_\|}\frac{\partial\bar{\bm{u}}}{\partial\bar{\mu}}\cdot\hat{\bm{b}}\times\left(\frac{\partial}{\partial t}+\bar{\bm{u}}\cdot\bar{\bm{\nabla}}+U\bar{\nabla}_\|\right)\bar{\bm{u}}\\&
-\frac{1}{\bar{\Omega}^*_\|}\frac{\partial\bar{\bm{u}}}{\partial\bar{\mu}}\cdot\left(\hat{\bm{b}}\times\bar{\bm{\nabla}}\langle\phi\rangle-\Omega\bar{\bm{u}}\right).
\end{align*}
Eq. \eqref{el} takes its simplest form when $\bm{u}$ is chosen to be the $\rm E\times B$ drift velocity associated with the $\theta$-independent potential that appears in Eq. \eqref{gc}.
\section{Poisson equation}\label{appendix_poisson}
The variation with respect to $\phi$ of the gyrocentre system Lagragian up to second order is\begin{eqnarray*}
(\delta L)_\phi&=&-\int{\rm d}^6\bar{Z}\bar{F}\{\delta[\langle\phi\rangle-\tfrac{1}{2}\Omega^{-2}\bar{\bm{\nabla}}\tilde{\Phi}\times\hat{\bm{b}}\cdot\bar{\bm{\nabla}}\tilde{\phi}\\&&
-\tfrac{1}{2}\Omega^{-1}\langle\tilde{\phi}^2\rangle_{,\bar{\mu}}+\Omega^{-1}\bar{\bm{\nabla}}_\perp\langle\phi\rangle\cdot(\Omega^{-1}\bar{\bm{\nabla}}_\perp\langle\phi\rangle\\&&
-\dot{\bar{\bm{R}}}\times\hat{\bm{b}})]\}_\phi\\&
=&-\int{\rm d}^6\bar{Z}\bar{F}(\{\langle\phi+\delta\phi\rangle\\&&
-\tfrac{1}{2}\Omega^{-2}\bar{\bm{\nabla}}(\tilde{\Phi}+\delta\tilde{\Phi})\times\hat{\bm{b}}\cdot\bar{\bm{\nabla}}(\tilde{\phi}+\delta\tilde{\phi})\\&&
-\tfrac{1}{2}\Omega^{-1}\langle(\tilde{\phi}+\delta\tilde{\phi})^2\rangle_{,\bar{\mu}}\\&&
+\Omega^{-1}\bar{\bm{\nabla}}_\perp\langle\phi+\delta\phi\rangle\cdot[\Omega^{-1}\bar{\bm{\nabla}}_\perp\langle\phi+\delta\phi\rangle-\dot{\bar{\bm{R}}}\times\hat{\bm{b}}]\}\\&&
-[\langle\phi\rangle-\tfrac{1}{2}\Omega^{-2}\bar{\bm{\nabla}}\tilde{\Phi}\times\hat{\bm{b}}\cdot\bar{\bm{\nabla}}\tilde{\phi}\\&&
-\tfrac{1}{2}\Omega^{-1}\langle\tilde{\phi}^2\rangle_{,\bar{\mu}}+\Omega^{-1}\bar{\bm{\nabla}}_\perp\langle\phi\rangle\cdot(\Omega^{-1}\bar{\bm{\nabla}}_\perp\langle\phi\rangle\\&&
-\dot{\bar{\bm{R}}}\times\hat{\bm{b}})])\\&
=&-\int{\rm d}^6\bar{Z}\bar{F}[\langle\delta\phi\rangle-\Omega^{-2}\langle\bar{\bm{\nabla}}\tilde{\Phi}\times\hat{\bm{b}}\cdot\bar{\bm{\nabla}}\delta\phi\rangle\\&&
-\Omega^{-1}\langle\tilde{\phi}\delta\phi\rangle_{,\bar{\mu}}+\Omega^{-1}\bar{\bm{\nabla}}_\perp\langle\delta\phi\rangle\cdot(\Omega^{-1}\bar{\bm{\nabla}}_\perp\langle\phi\rangle\\&&
-\dot{\bar{\bm{R}}}\times\hat{\bm{b}})+\Omega^{-2}\bar{\bm{\nabla}}_\perp\langle\phi\rangle\cdot\bar{\bm{\nabla}}_\perp\langle\delta\phi\rangle],\end{eqnarray*}
from which we obtain the Euler-Lagrange equation for $\phi$ \eqref{dl}.

Using an alternative form for $\bar{\Gamma}_2$ \eqref{g2} and $J_{\bar{Z}\to z}=\bar{\Omega}^*_\|$, the Euler-Lagrange equation for $\phi$ up to first order is\begin{equation}\begin{aligned}
0=&\Omega\int{\rm d}^6\bar{Z}\delta(\bar{\bm{R}}+\bar{\bm{\rho}}-\bm{r})[(1+\Omega^{-2}\bar{\bm{\nabla}}\tilde{\Phi}\times\hat{\bm{b}}\cdot\bar{\bm{\nabla}}\\&
+\Omega^{-1}\tilde{\phi}\partial_{\bar{\mu}})\bar{F}'+\Omega^{-2}\bar{\nabla}_\perp^2\langle\phi\rangle\bar{F}'-\Omega^{-1}\bar{\bm{\rho}}\cdot(\bar{F}'\bar{\bm{\nabla}}\langle\phi\rangle)_{,\bar{\mu}}].\end{aligned}\label{p0}\end{equation}
Using the guiding-centre Jacobian up to first order $J_{Z\to z}=\Omega^*_\|+\bm{\rho}\cdot\bm{\Omega}\times\bm{u}_{,\mu}$ and the action of the Lie transform on scalars up to first order $\mathsf{T}\bar{F}'=(1+g^i_1\partial_i)\bar{F}'$, an evaluation of Eq. \eqref{nd1} up to first order yields Eq. \eqref{p0}. In other words, we obtain equivalent Poisson equations up to first order using either a variational or direct method.

We will now consider uniform $\bar{F}'$. Using $\bar{\bm{\nabla}}\langle\phi\rangle=-\int{\rm d}^3k\langle\bm{E}\rangle(\bm{k},\bar{\mu})e^{i\bm{k}\cdot\bar{\bm{R}}}$, the last two terms in Eq. \eqref{p0} are\begin{eqnarray*}
2\pi i\int{\rm d}\bar{U}{\rm d}\bar{\mu}{\rm d}^3k\{[\bar{\rho}J_1(k_\perp\bar{\rho})]_{,\bar{\mu}}&&\\
-k_\perp\Omega^{-1}J_0(k_\perp\bar{\rho})\}\langle\bm{E}\rangle e^{i\bm{k}\cdot\bm{r}}\bar{F}'&=&0.\end{eqnarray*}
In other words, in the weak-flow limit and for uniform $\bar{F}'$, the weak- and strong-flow Poisson equations up to first order are identical,\[
0=\Omega\int{\rm d}^6\bar{Z}\delta(\bar{\bm{R}}+\bar{\bm{\rho}}-\bm{r})(1+\Omega^{-1}\tilde{\phi}\partial_{\bar{\mu}})\bar{F}',\]
where, for uniform $\bar{F}'$, the second weak-flow polarisation density term does not appear.
\begin{acknowledgements}
This paper was sponsored in part by EPSRC grant EP/D062837/1. Computational facilities were provided by the MidPlus Regional Centre of Excellence for Computational Science, Engineering and Mathemtatics, under EPSRC grant EP/K000128/1.\end{acknowledgements}
\bibliography{_}\end{document}